\pdfoutput=1

\documentclass[11pt,twoside,a4paper,cmspaper,final,collab]{cms-tdr}
\def\svnVersion{329659}\def\cmsCernNo{2013-037}\def\cmsCernDate{\today}\def\cmsMessage{Published in the Journal of High Energy Physics as \href{http://dx.doi.org/10.1007/JHEP02(2016)156}{\doi{10.1007/JHEP02(2016)156.}}}

\begin{document}\cmsNoteHeader{HIN-14-016}

\hyphenation{had-ron-i-za-tion}
\hyphenation{cal-or-i-me-ter}
\hyphenation{de-vices}
\RCS$Revision: 329462 $
\RCS$HeadURL: svn+ssh://svn.cern.ch/reps/tdr2/papers/HIN-14-016/trunk/HIN-14-016.tex $
\RCS$Id: HIN-14-016.tex 329462 2016-03-01 23:09:00Z htrauger $

\newcommand {\sqrtsNN}  {\ensuremath{\sqrt{s_{_\mathrm{NN}}}}}
\providecommand{\PYHYD}{\textsc{pythia+hydjet}\xspace}
\providecommand{\HYDJET}{\textsc{hydjet}\xspace}
\hyphenation{dia-me-ter}
\hyphenation{pseudo-rapidity}

\newlength\cmsFigWidth
\ifthenelse{\boolean{cms@external}}{\setlength\cmsFigWidth{0.85\columnwidth}}{\setlength\cmsFigWidth{0.4\textwidth}}
\ifthenelse{\boolean{cms@external}}{\providecommand{\cmsLeft}{top}}{\providecommand{\cmsLeft}{left}}
\ifthenelse{\boolean{cms@external}}{\providecommand{\cmsRight}{bottom}}{\providecommand{\cmsRight}{right}}
\cmsNoteHeader{HIN-14-016}
\title{Correlations between jets and charged particles in PbPb and pp collisions at $\sqrtsNN= 2.76$\TeV}

\date{\today}

\abstract{

The quark-gluon plasma is studied via medium-induced changes to correlations between jets and charged particles in PbPb collisions compared to pp reference data. This analysis uses data sets from PbPb and pp collisions with integrated luminosities of 166\mubinv and 5.3\pbinv, respectively, collected at $\sqrtsNN = 2.76$\TeV. The angular distributions of charged particles are studied as a function of relative pseudorapidity ($\Delta\eta$) and relative azimuthal angle ($\Delta\phi$) with respect to reconstructed jet directions. Charged particles are correlated with all jets with transverse momentum ($\pt$) above 120\GeV, and with the leading and subleading jets (the highest and second-highest in $\pt$, respectively) in a selection of back-to-back dijet events. Modifications in PbPb data relative to pp reference data are characterized as a function of PbPb collision centrality and charged particle $\pt$. A centrality-dependent excess of low-$\pt$ particles is present for all jets studied, and is most pronounced in the most central events. This excess of low-$\pt$ particles follows a Gaussian-like distribution around the jet axis, and extends to large relative angles of $\Delta\eta\approx 1$ and $\Delta\phi\approx 1$.

}

\hypersetup{%
pdfauthor={CMS Collaboration},%
pdftitle={Correlations between jets and charged particles in PbPb and pp collisions at sqrt(s[NN])= 2.76 TeV},%
pdfsubject={CMS},%
pdfkeywords={CMS, physics, heavy ions, jets}}

\maketitle
\section{Introduction}

The quark-gluon plasma (QGP) produced in ultra-relativistic heavy ion collisions may be \linebreak
probed \emph{in situ} via partons produced in the initial hard-scatterings, which carry high transverse momentum ($\pt$) compared to most of the particles in the event. In the QGP, partons are expected to suffer energy loss in the medium, a phenomenon known as ``jet quenching"~\cite{Bjorken:1982tu}.  This effect was discovered at RHIC via observables including suppression of high-$\pt$ particle production~\cite{Phenix2002} and charged particle correlations~\cite{Adler:2002xw}.  Jet quenching has been studied at the CERN LHC via high-$\pt$ particle suppression~\cite{Aamodt:2010jd, CMS:2012aa, Aad:2015wga}, and via the momentum balance of reconstructed back-to-back dijets.  In these latter studies, dijet transverse momentum balance was investigated in PbPb, pPb and pp collisions~\cite{Aad:2010bu, Chatrchyan:2012nia, Chatrchyan:2014hqa, Chatrchyan:2011sx}. Significant imbalance was found in central PbPb collisions, consistent with a pathlength-dependent energy loss in the QGP medium. In peripheral PbPb collisions and in pPb collisions, the dijet momentum balance is comparable to the one measured in pp collisions, thus confirming that the energy loss in central PbPb collisions is not due to initial state cold nuclear matter effects.  In the QGP, the interaction of the hard-scattered partons (and their fragmentation products) with the medium leads to a redistribution of energy carried by the produced particles. Comparing the charged-particle distributions from heavy ion data to the pp reference can help to differentiate between energy loss models and ultimately constrain the properties of the QGP~\cite{Qin:2010mn, He:2011pd, Young:2011qx, Ma:2013pha}.

Early RHIC studies of two-particle correlations involving a leading high-$\pt$ (8--15\GeV) particle did not find a significant modification in the distribution of the associated particles at small angles from the leading particle. A quenching effect was found in the distribution of particles opposite in azimuth to the leading particle, observed as a reduction in the associated yield~\cite{Star2003, Star2005, RhicDijets, RhicJetHadron, Adare:2007vu}.  These results could be interpreted as an in-medium energy loss (in which associated particles fully thermalize and do not retain correlation to the jet direction) followed by a vacuum-like fragmentation of the remaining jet~\cite{Majumder:2010qh}.  Compared to the capabilities of the LHC, these studies at RHIC are significantly limited by the lower production rates for hard probes.  Subsequent high-precision measurements at the LHC~\cite{Chatrchyan:2012gw,Chatrchyan:2013kwa,CMS:2012vba} have shown that the detailed jet structures within a jet cone radius of 0.3 are modified by the medium in terms of both $\pt$ and angular distributions. However, these observed in-cone changes only explain a small fraction of the dijet momentum imbalance, indicating that a large amount of energy is radiated outside of the jet cone or transferred to particles with very low momentum.  Direct measurements of energy redistribution between event hemispheres containing subleading and leading jets were made by CMS via the ``missing-$\pt$" observable~\cite{Chatrchyan:2011sx,mpt}. These studies found the overall energy flow to be modified in PbPb collisions out to large radial distances from the dijet axis, and various theoretical models have since attempted to describe the result~\cite{Apolinario:2012cg, Ayala:2015jaa, Blaizot:2014ula}.  Extending measurements of jet structure modifications to similarly large angles is crucial to properly constrain the energy loss mechanism.

In this paper, we use 166\mubinv of PbPb collisions taken during the 2011 LHC heavy ion run at a nucleon-nucleon center-of-mass energy of $\sqrt{s_{NN}} = 2.76$\TeV.  For the reference measurement, we use pp data taken in 2013 at the same center-of-mass energy corresponding to an integrated luminosity of 5.3\pbinv.  Two-dimensional angular correlations were previously studied in CMS for pairs of charged particles~\cite{Chatrchyan:2011eka}.  In the present analysis, this technique is applied to correlate jets with charged particles.  For each charged particle and reconstructed jet, $\pt$ and pseudorapidity ($\eta$) are measured with respect to the beam axis, and azimuthal angle ($\phi$) is measured in the transverse plane.  From these measurements, the relative pseudorapidity ($\Delta\eta = \eta_{\text{track}} -\eta_{\text{jet}}$) and relative azimuth ($\Delta\phi = \phi_{\text{track}} -\phi_{\text{jet}}$) between jets and charged-particle tracks is determined.  These relative angles are used to construct two-dimensional $\Delta\eta$--$\Delta\phi$ charged particle density distributions, which we will refer to as ''jet-track correlations''.  The jet-track correlations  are then studied as a function of centrality (defined as a percentile of the total inelastic cross section, with 0\% indicating collisions with impact parameter $b = 0$) and charged-particle transverse momentum ($\pt^\text{trk}$).

In order to extend these measurements to low $\pt^\text{trk}$, where soft particles resulting from energy loss mechanisms such as gluon radiation are expected to appear, an analysis must carefully handle both large combinatorial backgrounds typical for the heavy ion environment, and long-range correlations arising from hydrodynamic expansion~\cite{Alver:2010gr}. Taking advantage of the kinematic reach of hard probes and the availability of detailed characterization of the event bulk properties, the CMS detector permits the statistical separation of the medium-related modifications of jet-track correlations from the long-range hydrodynamic background.  Using this technique, this study captures jet-related energy flow both inside and outside of the jet cone, extending measurements of intrinsic jet properties to large relative angles in $\Delta\eta$ and $\Delta\phi$.

\section{The CMS detector}

The central feature of the CMS apparatus is a superconducting solenoid with an internal diameter of 6\unit{m}, providing an axial uniform magnetic field of 3.8 T.
Muons are measured in gas-ionization detectors embedded in the steel flux-return yoke outside the solenoid.  Within the solenoid volume are a silicon pixel and strip tracker, a lead tungstate crystal electromagnetic calorimeter (ECAL), and a brass and scintillator hadron calorimeter (HCAL), each composed of a barrel and two endcap sections.  Extensive hadronic forward (HF) steel and quartz fiber calorimetry complements the barrel and endcap detectors, providing coverage to $\abs{\eta}<5$. In this analysis, the collision centrality is determined using the total sum of transverse energy ($\ET$) from calorimeter towers in the HF region (covering \mbox{$2.9<\abs{\eta}<5.2$}). For the forward region $2.9<\abs{ \eta } < 5.0$ relevant for collision centrality determination, HF fibers are bundled in towers with widths of $0.175 \times 0.175$ ($\Delta\eta \times \Delta\phi$)~\cite{Chatrchyan:2011ds}.  The $\ET$ distribution is used to divide the event sample into bins, each representing 0.5\% of the total nucleus-nucleus hadronic interaction cross section.  A detailed description of centrality determination can be found in Ref.~\cite{Chatrchyan:2012nia}.

Jet reconstruction for this analysis relies on calorimeter information from the ECAL and HCAL.  For the central region $\abs{ \eta } < 1.6$ in which jets are selected for this analysis, the HCAL cells have widths of 0.087 in both $\eta$ and $\phi$. In the $\eta$-$\phi$ plane, and for $\abs{\eta} < 1.48$, the HCAL cells map on to $5 \times 5$ ECAL crystals arrays to form calorimeter towers projecting radially outwards from close to the nominal interaction point. The barrel section of the ECAL has an energy resolution of 1\%--2.5\%, while the endcaps have an energy resolution of 2.5--4\%~\cite{CMS:EGM-14-001}. Within each tower, the energy deposits in ECAL and HCAL cells are summed to define the calorimeter tower energies, subsequently used to provide the energies and directions of hadronic jets. When combining information from the entire detector, the jet energy resolution amounts typically to 15\% at 10\GeV, 8\% at 100\GeV, and 4\% at 1\TeV, to be compared to about 40\%, 12\%, and 5\% obtained when the ECAL and HCAL calorimeters alone are used~\cite{Chatrchyan:2011ds}.

Accurate particle tracking is critical for measurements of charged-hadron yields.  The CMS silicon tracker measures charged particles within the region $\abs{\eta}< 2.5$. For particles of $1 < \pt < 10$\GeV and $\abs{\eta} < 1.4$, the track resolutions are typically 1.5\% in \pt and 25--90 (45--150)\mum in the transverse (longitudinal) impact parameter with respect to the collision vertex~\cite{TRK-11-001}.  A detailed description of the CMS detector, together with a definition of the coordinate system used and the relevant kinematic variables, can be found in Ref.~\cite{CMS_Detector}.

\section{Jet and track reconstruction and corrections}
\label{sec:Jetreco}

For both pp and PbPb collisions, jet reconstruction is performed with the anti-\kt algorithm, encoded in the \textsc{FastJet} framework~\cite{bib_antikt, fastjet1}.  Following previous CMS studies~\cite{Chatrchyan:2011sx, Chatrchyan:2012gw,Chatrchyan:2013kwa,CMS:2012vba}, a narrow jet reconstruction distance parameter, $R = 0.3$, is chosen due to the large underlying event in heavy ion collisions. Jet $\pt$ and direction in $\eta$ and $\phi$ are determined based on iterative clustering of energy deposits in the CMS calorimeters. For PbPb collisions, the CMS algorithm ``HF/Voronoi'' is used to subtract the heavy ion underlying event~\cite{CMS-DP-2013-018}.  This algorithm estimates the underlying event contribution to the $\ET$ in each calorimeter tower by performing a singular value decomposition of the particle distributions.  The average $\ET$, as a function of $\eta$ and $\phi$, is subtracted from each calorimeter tower, and then the energy is redistributed between neighboring calorimeter towers so that no tower contains non-physical negative $\ET$.  For pp collisions no underlying event subtraction is employed, as the effect of the underlying event on the jet energy is small relative to the jet energy scale (JES) uncertainty.

Monte Carlo (MC) event generators have been used for evaluation of the jet and track reconstruction performance, in particular for determining the tracking efficiency as well as the jet energy response and resolution. Jet events are generated with \PYTHIA~\cite{bib_pythia} (version 6.423, tune Z2~\cite{Field}). These generated \PYTHIA events are propagated through the CMS detector using the \GEANTfour package~\cite{GEANT4} to simulate
the detector response. In order to account for the influence of the underlying PbPb events, the \PYTHIA events are embedded into fully simulated PbPb events that are generated by \HYDJET~\cite{Lokhtin:2005px} (version 1.8), which is tuned to reproduce the total particle multiplicities, charged-hadron spectra, and elliptic flow at all centralities. The embedding is performed by mixing the simulated digital signal information from \PYTHIA and \HYDJET, hereafter referred to as \PYHYD. No simulation of jet quenching is applied in this \PYHYD simulation.  These events are then propagated through the same reconstruction and analysis procedures used for data events.  The JES is established using \PYTHIA and \PYHYD events in bins of event centrality as a function of $\pt$, $\eta$, and number of charged particles inside the jet cone.  For studies of pp data and \PYTHIA simulation, charged particles are reconstructed using the same iterative method~\cite{TRK-11-001} as in the previous CMS analyses of pp collisions. For PbPb data and \PYHYD MC, an iterative charged-particle reconstruction similar to that of earlier heavy ion analyses~\cite{CMS:2012aa,Chatrchyan:2013kwa} is employed, as described in detail in Ref.~\cite{mpt}.

\section{Data samples and triggers}
\label{sec:evtsel}

The first level (L1) of the CMS trigger system, composed of custom hardware processors, uses information from the calorimeters and muon detectors to select the most interesting events in a fixed time interval of less than 4\mus. The high-level trigger (HLT) processor farm further decreases the event rate from around 100\unit{kHz} to less than 1\unit{kHz} before data storage.  The events for this analysis were selected using an HLT that selects all events containing at least one calorimeter jet with $\pt> 80$\GeV.  The HLT is fully efficient for events containing offline reconstructed jets with $\pt > 120$\GeV. In order to suppress noise due to noncollision sources such as cosmic rays and beam backgrounds, the events used in this analysis were further required to satisfy offline selection criteria as documented in Refs.~\cite{Chatrchyan:2012nia,Chatrchyan:2012gt}. These criteria include selecting only events with a reconstructed vertex including at least two tracks and a $z$ position within 15 cm of the detector center, and requiring energy deposits in at least 3 forward calorimeter towers on either side of the interaction point.

The offline selection of events begins with jets reconstructed as described in Section~\ref{sec:Jetreco}.  To study the jet-track correlations, the events are then categorized into two samples:  an inclusive selection of high-$\pt$ jets and a selection of back-to-back dijet events.  For the inclusive sample, jets are required to have $\pt > 120$\GeV and to fall within $\abs{\eta} < 1.6$, with multiple jets from the same event permitted in this inclusive jet sample.  These inclusive selection criteria match previous CMS studies~\cite{CMS:2012vba, Chatrchyan:2013kwa} that measured jet fragmentation functions and jet shapes within the jet cone ($\Delta R < 0.3$), and allow this analysis to extend comparable measurements to large angles from the jet axis.  We then separately select a dijet sample using criteria matched to those of a previous analysis that explores dijet energy balance~\cite{Chatrchyan:2011sx}.  In this dijet selection, events are first required to contain a highest $\pt$ (leading) calorimeter jet in the range of $\abs{\eta} < 2$, with a corrected jet $\pt> 120$\GeV and a next-highest $\pt$ (subleading) jet of $\pt > 50$\GeV, also in $\abs{\eta}<2$.  The azimuthal angle between the leading and subleading jets is required to be at least $5\pi/6$. No explicit requirement is made either on the presence or absence of a third jet in the event.  To ensure stable jet reconstruction performance, only events in which both leading and subleading jets fall within $\abs{\eta}<1.6$ are included in the final data sample.

\section{Jet-track angular correlations}

Charged tracks in the event with $\pt^\text{trk}$ above 1\GeV are used to construct two-dimensional $\Delta\eta $--$\Delta\phi$ correlations with respect to the individually measured jet axis for inclusive jets and for leading and subleading jets in dijet events.  The jet-track correlations are constructed according to the procedure established in Ref.~\cite{HIN13002}, for the following bins in $\pt^\text{trk}$:  1--2, 2--3, 3--4, and 4--8\GeV.  This work does not attempt to construct correlations below 1\GeV, where the jet-related signal is very small compared to the combinatorial and long-range correlated background, or for $\pt^\text{trk} > 8$\GeV where the statistical power becomes limited.  The correlations are corrected for tracking efficiency and misreconstruction on a per-track basis, using an efficiency parametrization defined as a function of centrality, $\pt^\text{trk}$, $\eta$, $\phi$, and radial distance from the nearest jet with $\pt > 50$\GeV~\cite{mpt}.

 Correlations are formed by measuring angular distances to the inclusive, leading and subleading jet axes for each $\pt^\text{trk}$ range.  The signal pair distribution, $S(\Delta\eta,\Delta\phi)$, represents the per-track efficiency-corrected yield of jet-track pairs $N^{\rm same}$ from the same event normalized by the total number of jets:
 \begin{linenomath}
  \begin{equation}
  \label{eq:signal}
  S(\Delta\eta,\Delta\phi) = \frac{1}{N_\text{jets}}\frac{\rd^{2}N^\text{same}}{\rd\Delta\eta\, \rd\Delta\phi}.\
    \end{equation}
    \end{linenomath}

To correct for pair acceptance effects, we use the mixed event technique~\cite{ppridge,Chatrchyan:2011eka,pPbridge} to determine the geometrical $\Delta\eta$--$\Delta\phi$ shape that arises from selecting jets and tracks from within our respective acceptances of $\abs{\eta_\text{ jet}}<1.6$ and $\abs{\eta_\text{track}}<2.4$. In this technique, a mixed event distribution, $ME(\Delta\eta,\Delta\phi)$, is constructed by correlating the reconstructed jet axis direction from a selected signal event to tracks from events in a minimum bias data sample.  For each signal event, 40 minimum bias events are selected to have a similar vertex position (within 1\unit{cm}) and event centrality (within 2.5\%) to the jet event. Mixed event correlations are corrected for tracking efficiency and misreconstruction on a per-track basis applying the same efficiency parametrization used to correct signal correlations. The distribution of such mixed event jet-track pairs $N^\text{mix}$ is denoted:
\begin{linenomath}
  \begin{equation}
  \label{eq:mixed}
  ME(\Delta\eta,\Delta\phi) = \frac{1}{N_\text{jets}}\frac{\rd^{2}N^\text{mix}}{\rd\Delta\eta\, \rd\Delta\phi}.\
  \end{equation}
  \end{linenomath}

This distribution $ME(\Delta\eta,\Delta\phi)$ is normalized to unity at ($\Delta\eta=0,\Delta\phi=0$), where jets and tracks are close together and therefore have full pair acceptance.  Correlations are corrected for \linebreak
pair acceptance effects by dividing them by the normalized mixed event distribution \linebreak
$ME(\Delta\eta,\Delta\phi)/ME(0,0)$. The resulting yield of associated tracks per jet is defined as:
\begin{linenomath}
  \begin{equation}
  \label{2pcorr_incl}
  \frac{1}{N_\text{jets}}\frac{\rd^{2}N}{\rd\Delta\eta\, \rd\Delta\phi}
  = ME(0,0)\,\frac{S(\Delta\eta,\Delta\phi)}{ME(\Delta\eta,\Delta\phi).}
  \end{equation}
   \end{linenomath}

\begin{figure}[h!]
\centering
\includegraphics[width=0.99\textwidth]{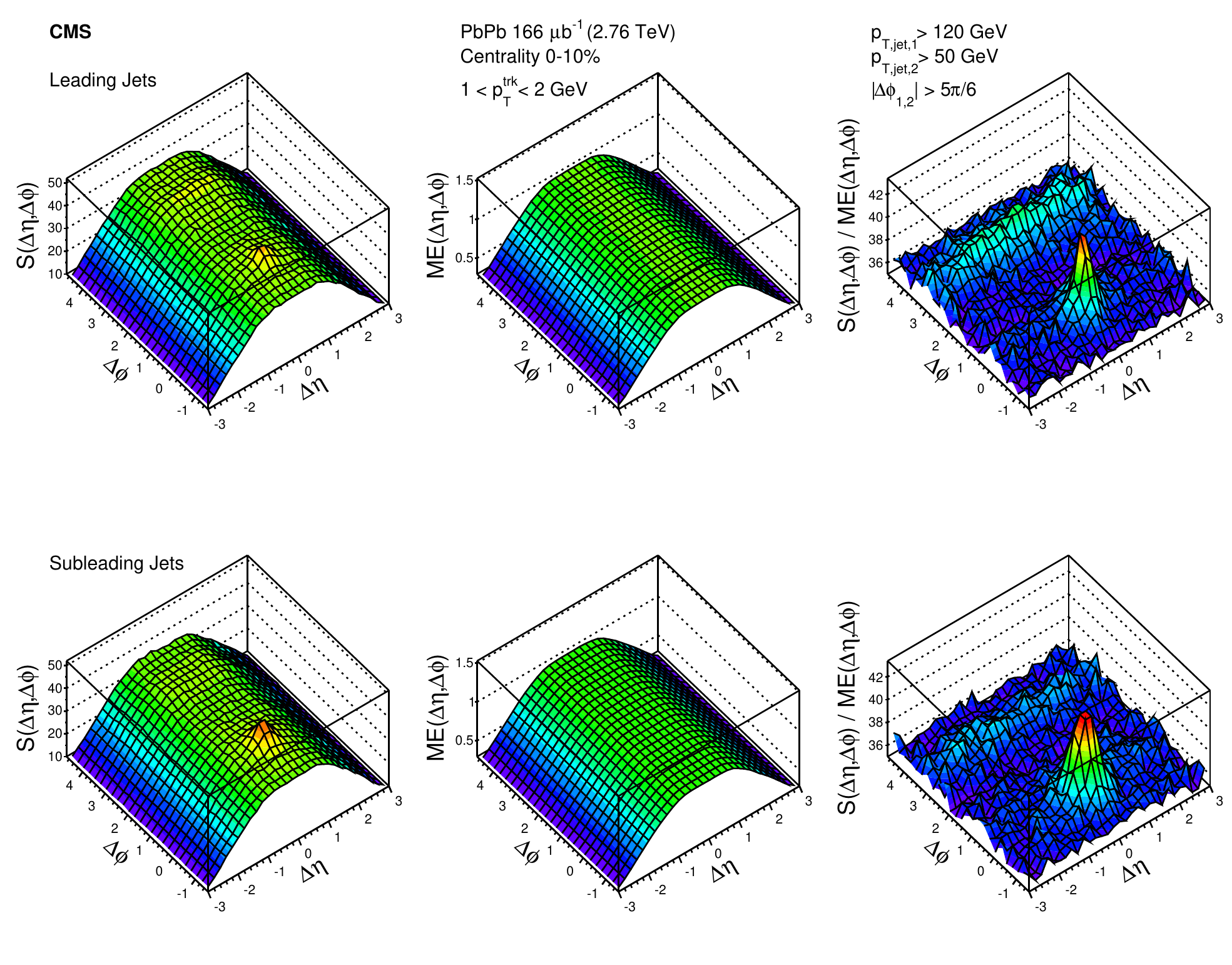}

\caption{Jet-track correlation signal shape $S(\Delta\eta,\Delta\phi)$ for tracks with 1 $ < \pt^\text{trk} < $ 2\GeV in 0--10\% central events (left), and corresponding mixed event shape $ME(\Delta\eta,\Delta\phi)$ for the same centrality and $\pt^\text{trk}$ bin (center). Their ratio gives the acceptance-corrected yield (right).  The top row shows the correlation between leading jets (with $p_{\mathrm{T},\text{jet1}}> $ 120\GeV) and all tracks, while the bottom row shows the correlation between subleading jets (with $p_{\mathrm{T},\text{jet2}}>$ 50\GeV) and all tracks. }
\label{fig:Correlation_0_10}
\end{figure}

This process is illustrated in Fig.~\ref{fig:Correlation_0_10}: the raw correlations to the leading and subleading jets in dijet events are shown on the left for the lowest $\pt^\text{trk}$ bin (1--2\GeV) and 0--10\% centrality range. The jet-like peak at $(\Delta\eta,\Delta\phi) = (0,0)$ is visible about both the leading and the subleading jets despite the high background levels in these most central events.  An away-side peak at $(\Delta\eta,\Delta\phi) = (0,\pi)$ is also visible in both leading and subleading jet correlations, corresponding in the leading jet correlation to the $\Delta\eta$-smeared subleading jet peak, and likewise corresponding to the $\Delta\eta$-smeared leading jet peak in the subleading jet correlation.  The middle panel shows the shape of the pair acceptance correction determined using the mixed event technique. Finally, on the right, we present the acceptance-corrected correlations for the same $\pt^\text{trk}$ bin before the subtraction of long-range correlation terms.

\begin{figure}[ht!]
\centering
\includegraphics[width=0.99\textwidth]{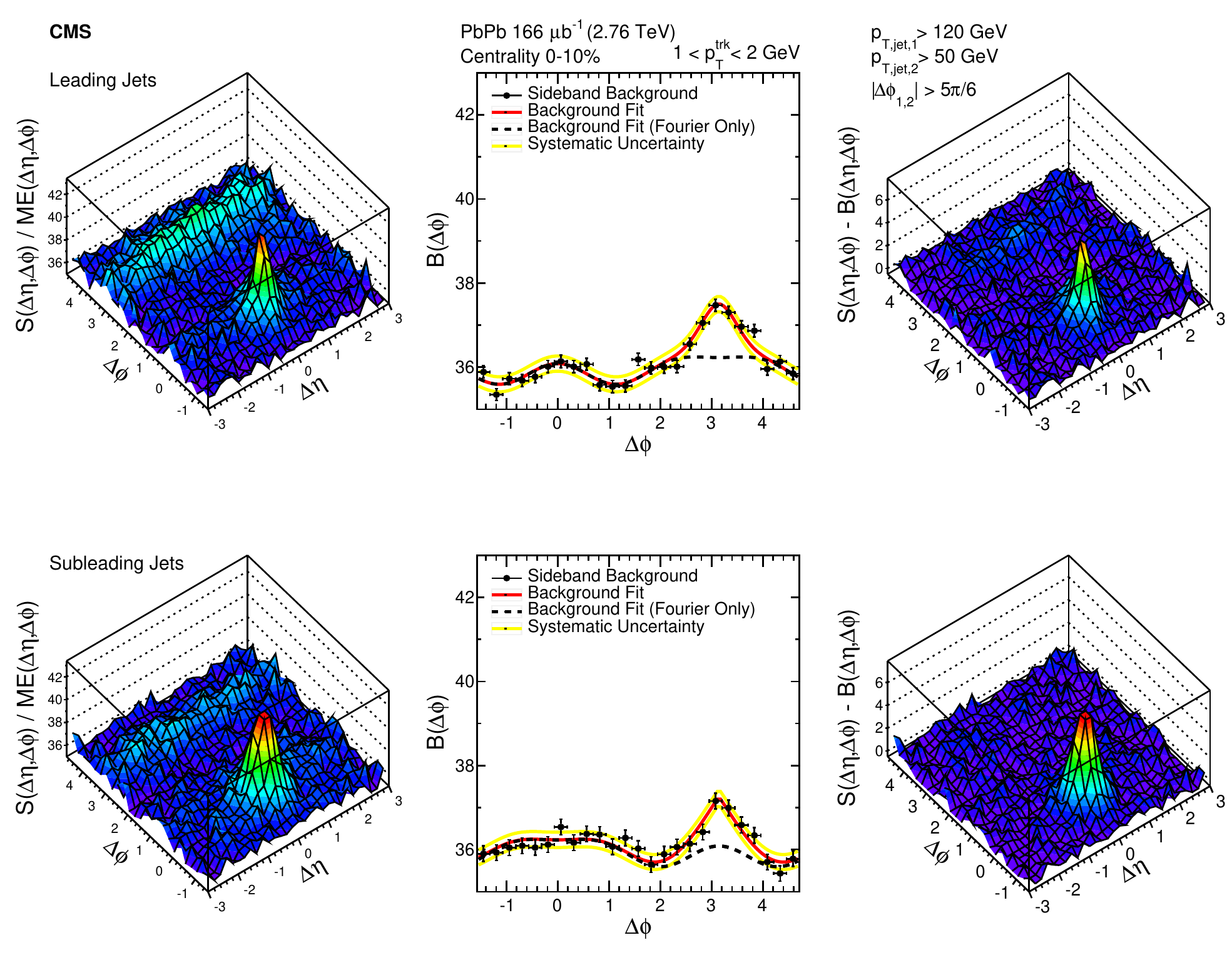}

\caption{Acceptance-corrected 2D jet-track correlation yield (left) is projected over the range $1.5<\abs{\Delta\eta}<3.0$, producing a 1D background distribution (center).  The fit to this distribution (indicated with a red dark line) is subtracted from the total yield to obtain the 2D background-subtracted yield shown on the right (for tracks with 1$ < \pt^\text{trk} < 2\GeV$).  The black dashed line shows the background level and Fourier flow harmonic components of the fit only, excluding the away-side peak.  Yellow lines in the $B(\Delta\phi)$ plot (middle panel) indicate the systematic uncertainty assigned to the background subtraction.}
\label{fig:Bkg_subtraction_illustration_0_10}
\end{figure}

To subtract the random combinatorial backgrounds and long-range correlations (dominated by hydrodynamic flow in PbPb and momentum conservation constraints in pp events), we employ a sideband subtraction technique in which these backgrounds are approximated by the measured two--dimensional correlations in the range $1.5<\abs{\Delta\eta}<3.0$.   Based on a CMS study that shows no appreciable variation of the elliptic flow for charged particles with $\pt^\text{trk} > 1$\GeV in the $\Delta\eta$ interval of $\pm$3.0 relevant for the present analysis~\cite{v2_HIN_11_012}, the Fourier harmonics are assumed to be constant in $\Delta\eta$. This background distribution in relative azimuthal angle (integrated over $1.5<\abs{\Delta\eta}<3.0$) is then fitted with a function modeling harmonic flow plus a term to capture the (Gaussian or sharper) peak at $\Delta\phi = \pi$ associated with the (smeared) jet opposite to the jet under study:

\begin{linenomath}
\begin{equation}
B(\Delta\phi) = B_{0}(1+2V_{1}\cos{(\Delta\phi)}+2V_{2}\cos{(2\Delta\phi)}+2V_{3}\cos{(3\Delta\phi)})+A_\mathrm{ AS}\exp{\bigg[-\bigg(\frac{\abs{\Delta\phi-\pi}}{\alpha}\bigg)^{\beta} \bigg] },
\end{equation}
\end{linenomath}

where $B_{0}$ is the overall background level; $V_{1}$, $V_{2}$, and $V_{3}$ are Fourier coefficients modeling harmonic flow; and $A_\mathrm{AS}$, $\alpha$, and $\beta$ are respectively the magnitude, width, and shape parameters of the away-side peak.  We find that at low $\pt^\text{trk}$ the long-range azimuthal sideband distributions are exhausted by the first three Fourier coefficients ($V_{1}$, $V_{2}$, $V_{3}$), while at high $\pt^\text{trk}$, $V_{1}$ and $V_{2}$ are sufficient to describe the background level within statistical uncertainties.  Figure~\ref{fig:Bkg_subtraction_illustration_0_10} illustrates the background subtraction process. The long-range contributions of the full 2D correlation (left) are estimated by the $\Delta \phi$ projection (shown in the middle panel) of this correlation over the range $1.5<\abs{\Delta \eta}<3.0$.  The fit to this $\Delta\phi$ distribution is propagated uniformly in $\Delta\eta$, and subtracted from the acceptance-corrected yield.  The short-range correlations remaining after this background subtraction are shown on the right panel, again for $\pt^\text{trk}= $ 1--2\GeV.

Jet-track correlations obtained from PbPb data are compared with those obtained from the pp reference data. To ensure that the kinematic range of the jets included in this comparison is the same, correlations are reweighted on a jet-by-jet basis so that the resulting jet $\pt$ spectrum matches that of PbPb data for a given centrality class.  Weighting factors are derived from the ratio of the normalized PbPb to pp jet spectra in bins of 10\GeV.  The reference pp jet $\pt$ spectrum is also smeared to account for jet energy resolution differences between  the PbPb and pp samples.  Reference correlations in $\Delta\eta$ and $\Delta\phi$ are then constructed and analyzed following the procedure described above for PbPb data.

\section{Corrections and systematic uncertainties}
\label{sec:systematics}

An analysis of \PYTHIA and \PYHYD MC simulated events is performed to evaluate and correct for the effects of two jet reconstruction biases on the measured correlated yield:  a bias toward the selection of jets with harder fragmentation, and a bias toward the selection of jets that coincide with upward fluctuations in the background.  The first correction addresses a jet fragmentation function (JFF) bias in which the jet energy is over-estimated for jets with hard fragmentation and under-estimated for those with soft fragmentation, resulting in a preferred selection of jets with harder fragmentation.  This bias affects pp and PbPb data similarly, and results in a reduction in the charged-particle correlated yield.  To correct for this effect in pp data, we compare correlations between reconstructed versus generated jets and generated particles in \PYTHIA simulations, and subtract the difference (reconstructed minus generated) from data correlations.  For the corresponding PbPb correction, we consider \PYTHIA jets embedded into and reconstructed within a \HYDJET-simulated environment, comparing correlations between generated versus reconstructed jets and the generated particles corresponding to the embedded \PYTHIA hard-scattering.  We note that this procedure also corrects for jet axis smearing in reconstruction, which is found to have no significant effect on the total integral of the correlation, and to affect the correlation shapes only within $\Delta\eta < 0.2$ and $\Delta\phi < 0.2$.  The magnitude of this correction (relative to the total correlated yield) ranges from 3 to 6\% in pp data, and from 3 to 7\% in PbPb data.  In \PYHYD, this JFF bias correction is found to be centrality-independent (and very similar to that for \PYTHIA), and is applied as a single correction for all centrality bins.  Maximum variations between centrality bins are used to evaluate the systematic uncertainty in this correction, which is found to be within 2\% of the correlated yield at low $\pt^\text{trk}$ and decreasing to zero at high $\pt^\text{trk}$.

The second correction evaluates and subtracts the measured charged-particle yield resulting from the preferential selection of jets that coincide with upward fluctuations in the background as detailed in Ref.~\cite{CMS:2012vba}.  The selections of inclusive and leading jets with  $\pt > 120$\GeV and subleading jets with $\pt > 50 $\GeV are sensitive to fluctuations in the background. Lower-energy jets that coincide with upward fluctuations in the background are included in the sample, while higher-energy jets that coincide with downward background fluctuations are excluded.  Because the inclusive and leading jet $\pt$ spectra are both steeply falling, the inclusion in the sample of a jet coinciding with an upward fluctuation in the background is much more common than the exclusion of a jet coinciding with a downward fluctuation, resulting in an excess of background tracks near the jet axis.  To quantify this effect, we performed the full analysis using a sample of \PYTHIA jets embedded into a \HYDJET heavy ion environment, and then extracted the correlated yield (with respect to reconstructed jets) comprised of particles originating from the \HYDJET background.  This correction was also checked with a data-driven technique using minimum bias PbPb events to confirm that \HYDJET appropriately reproduces fluctuations in the PbPb background, and the resulting upward bias in charged-particle correlated yields when these fluctuations contribute to the reconstructed jet energy. This background fluctuation bias strongly depends on event centrality and $\pt^\text{trk}$, with a magnitude of up to 24\% of the corrected signal for the lowest $\pt$ tracks in the most central collisions, decreasing to within 3\% for high-$\pt$ tracks and to a negligible contribution in peripheral collisions. This correlated yield due to the background fluctuation bias is subtracted to correct PbPb data, and half its magnitude is applied as a systematic uncertainty.

In addition to the systematic uncertainty associated with these two jet-reconstruction-related corrections, other sources of systematic uncertainty in this analysis include the JES determination, track reconstruction, and the procedures applied to correct for pair acceptance effects and subtract the uncorrelated and long-range backgrounds.  The correlated yield uncertainty associated with the JES is assessed by varying the inclusive and leading jet $\pt$ selection threshold up and down by 3\% (according to the JES uncertainty and also including differences in quark versus gluon JES~\cite{mpt}).  The resulting maximum variations in total correlated particle yield are found to be within 3\% in all cases, and we assign a 3\% systematic uncertainty to account for this effect.  The uncertainties of the $\pt$-dependent tracking efficiency and misidentified track corrections are found to be within 3--4\% in PbPb and pp collisions, and are independent of the centrality of the collisions. To account for the possible track reconstruction differences in data and simulation, a residual 5\% uncertainty is applied based on observed variations in corrected to initial track $\pt$ and $\eta$ spectra for different track quality selections~\cite{mpt}.

We evaluate pair acceptance uncertainties by considering differences in the background levels measured separately in each of the two sideband regions of our acceptance-corrected correlations ($-3.0<\Delta\eta<-1.5$ and $1.5<\Delta\eta<3.0$). This results in an uncertainty within the range of 5--9\%.  The overall systematic uncertainty due to background subtraction is calculated by varying all fit parameters up and down by their respective uncertainties and calculating the maximum resulting differences in background level, and by considering the deviation from the ''0'' level after background subtraction in the sideband region $1.5<\abs{\Delta\eta}<3.0$.  In more central events (0--10\%), the background subtraction uncertainty is found to be within 2--5\% for the lowest $\pt^\text{trk}$ bin where the background is most significant compared to the signal level.

All systematic uncertainties, evaluated as a function of $\pt^\text{trk}$ and event centrality are summarized in Table~\ref{tab:sys} as fractions of the total measured yield. The range of uncertainties listed presents the variation with track transverse momentum, with larger uncertainty values corresponding to  the lowest $\pt^\text{trk}$ bin (1--2\GeV) for all sources.  The systematic uncertainties from all seven sources are added in quadrature to obtain the total systematic uncertainty, which is quoted as a fraction of the total charged-particle yield associated with the jet under study.

\begin{table}[h!]
\centering
\topcaption{Systematic uncertainties in the measurement of the jet-track correlations in PbPb and pp collisions, as percentage of the total measured correlated yield. The numbers presented in this table summarize the range of values of systematic uncertainty (as a function of $\pt^\text{trk}$) for different centrality bins.}
\label{tab:sys}
\begin{tabular}{l|rrrr|r}
\hline
\multicolumn{1}{c|}{Source} & 0--10\% &  10--30\% & 30--50\% & 50--100\% &  pp \\
\hline

Background fluctuation bias             & 3--12\% & 2--7\% & 1--5\% & 0--1\% &  --- \\
Jet fragmentation function bias       & 0--2\% &    0--2\% &   0--2\% & 0--2\% &  0--2\% \\
Residual jet energy scale                                   & 3\% &    3\% &   3\% &    3\% &   3\% \\
Tracking efficiency uncertainty          & 4\% & 4\% & 4\% & 4\% &   3\%  \\
Residual track efficiency corr.              &    5\% &    5\% &    5\% &    5\% &    5\% \\
Pair acceptance corrections                  & 5--9\% & 5--9\% & 4--8\% & 2--6\% &  2--3\% \\
Background subtraction                       & 2--5\% & 2--5\% & 2--5\% & 2--5\% &  1--2\% \\

\hline
Total                                        & 9--17\% & 9--14\% & 8--13\% & 8--10\% & 7--8\% \\

\hline

\end{tabular}
\end{table}

\section{Results}
\label{sec:results}

\begin{figure}[h!]
\centering
\includegraphics[width=0.99\textwidth]{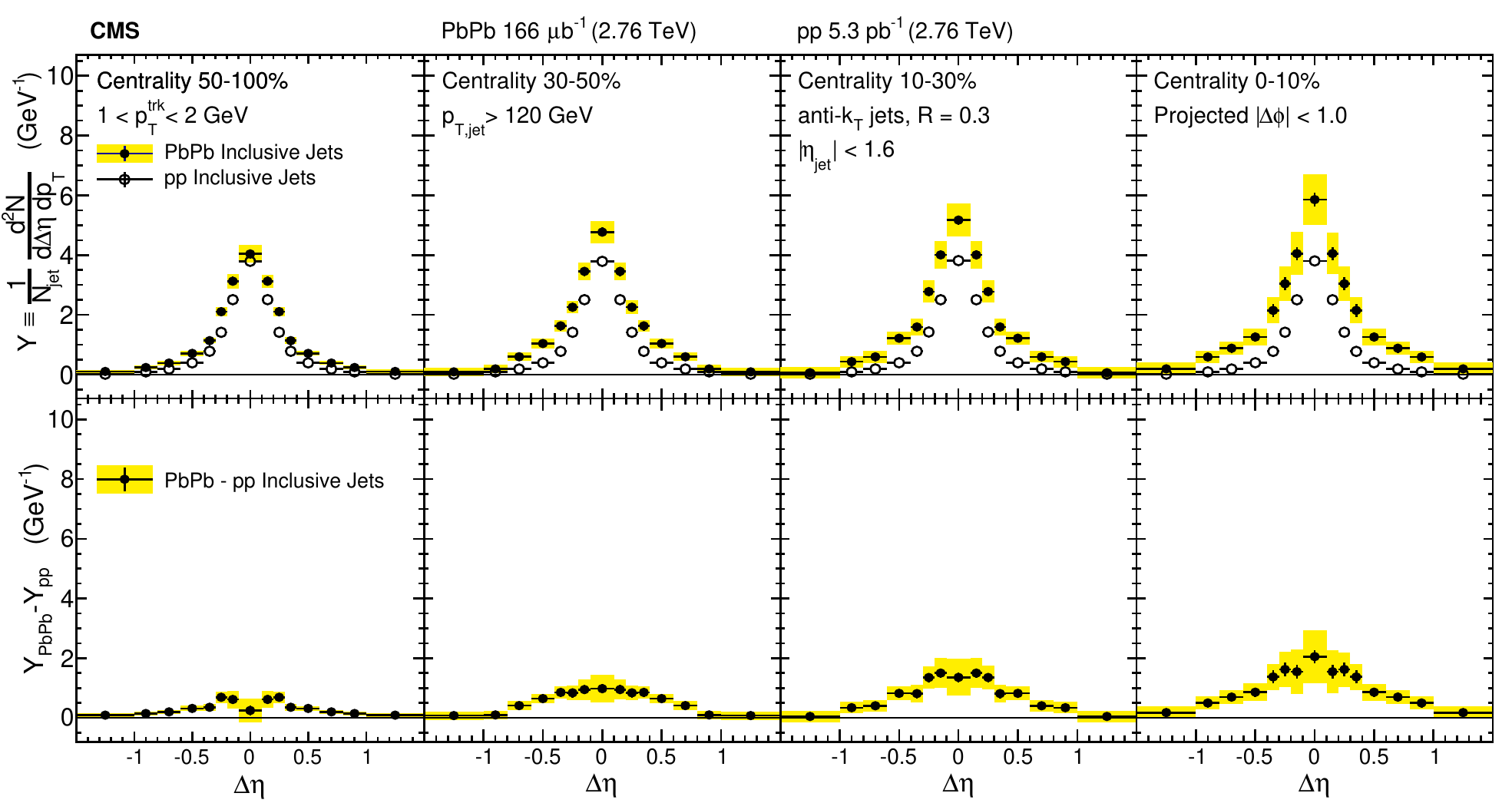}
\caption{Symmetrized $\Delta\eta$ distributions (projected over $\abs{\Delta\phi} < 1$) of background-subtracted particle yields correlated to PbPb and pp inclusive jets with $\pt>$ 120\GeV are shown in the top panels for tracks with 1 $ < p_{T}^\text{trk} < $ 2\GeV.  The difference in PbPb and pp per-jet yields is shown in the bottom panels. The total systematic uncertainties are shown as shaded boxes, and statistical uncertainties are shown as vertical bars (often smaller than the symbol size).}
\label{fig:Inclusive_dEta}
\end{figure}

\begin{figure}[h!]
\centering
\includegraphics[width=0.99\textwidth]{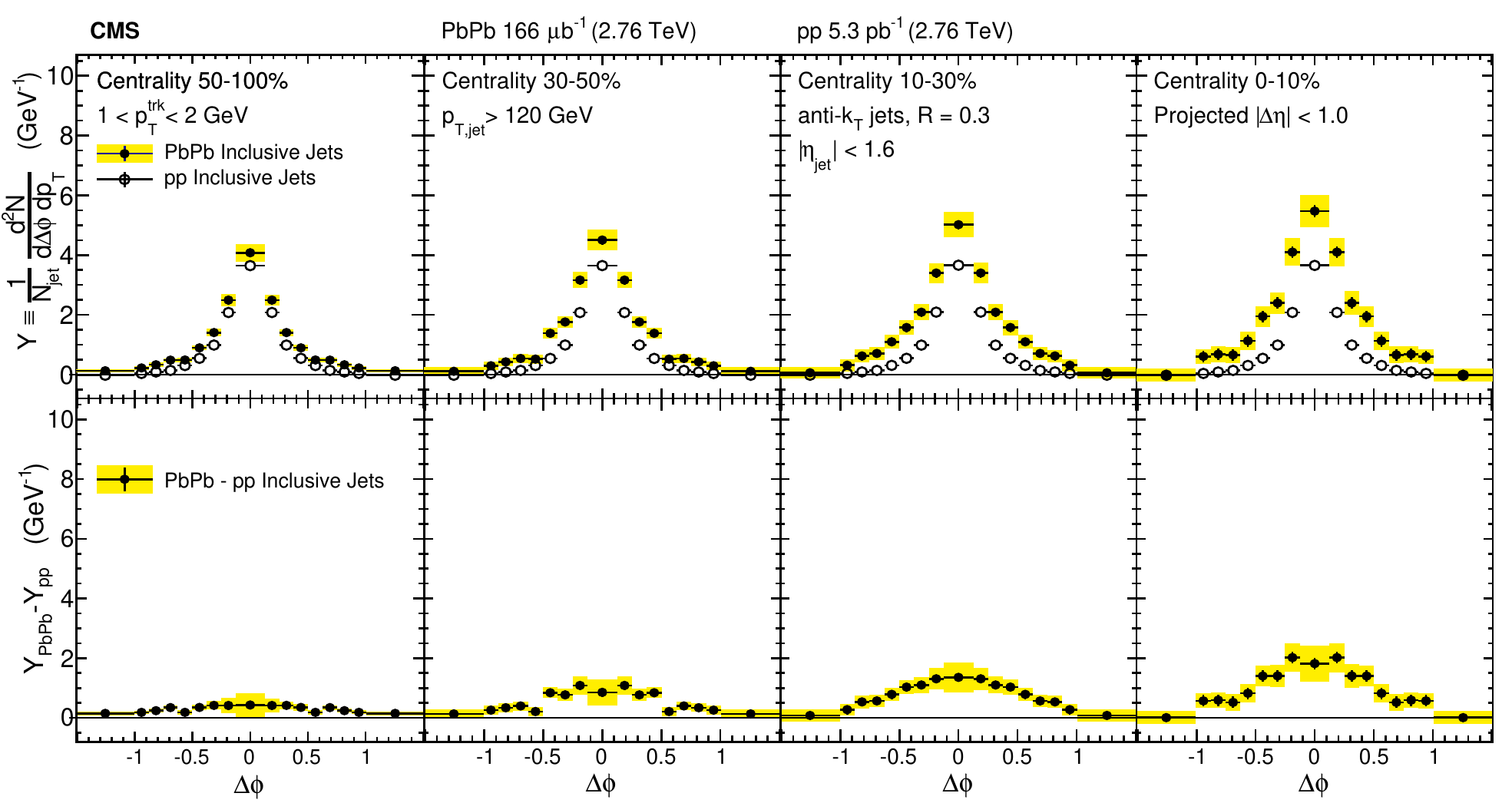}
\caption{Symmetrized $\Delta\phi$ distributions  (projected over $\abs{\Delta\eta} < 1$) of background-subtracted particle yields correlated to PbPb and pp inclusive jets with $\pt >$ 120\GeV are shown in the top panels for tracks with 1 $ < p_{T}^\text{trk} < $ 2\GeV.  The difference in PbPb and pp per-jet yields is shown in the bottom panels. The total systematic uncertainties are shown as shaded boxes, and statistical uncertainties are shown as vertical bars (often smaller than the symbol size).}
\label{fig:Inclusive_dPhi}
\end{figure}

In this analysis, jet-track correlations are studied differentially in centrality and $\pt^\text{trk}$. Correlations are projected in $\Delta\eta$ and $\Delta\phi$ to probe possible differences between azimuthal and pseudorapidity distributions. Figures~\ref{fig:Inclusive_dEta} and~\ref{fig:Inclusive_dPhi} show inclusive jet correlations projected on the $\Delta\eta$ (over $\abs{\Delta\phi}<1.0$) and $\Delta \phi$ (over $\abs{\Delta\eta}<1.0$) axes respectively for the lowest $\pt^\text{trk}$ selection. The upper panels of each figure present the centrality evolution of the correlations for inclusive jets with $\pt> 120$\GeV, together with a reference measurement from pp data at the same collision energy shown with open symbols.  To better visualize the PbPb to pp comparisons, the difference of the PbPb and pp correlation distributions is presented in the bottom panel for all centralities.  Correlations are symmetrized in $\Delta\eta$ and $\Delta\phi$ for clarity.

For the most peripheral events studied (centrality 50--100\%), the PbPb correlations at low transverse momentum, $1< \pt^\text{trk}<2$\GeV, show a very small excess (at most slightly larger than the uncertainties) relative to the pp reference data. This excess grows with collision centrality, with the most significant excess present in the most central collisions. The shape of this excess in the low-$\pt^{trk}$ per-jet particle yields is found to be similar in the $\Delta\eta$ and $\Delta\phi$ distributions, and in both dimensions exhibits a Gaussian-like shape that extends to large relative angles $\Delta\eta\approx 1$ and $\Delta\phi\approx 1$. We note that these results are consistent with previous CMS studies of jet-shape modifications~\cite{Chatrchyan:2013kwa} and fragmentation functions~\cite{CMS:2012vba} within the previously studied small $\Delta R < 0.3$ region, while extending measurements to individually study $\Delta\eta$ and $\Delta\phi$ distributions over the full range $\Delta\eta$ and $\Delta\phi < 1.5$.

\begin{figure}[hbtp]
\centering
\includegraphics[width=0.99\textwidth]{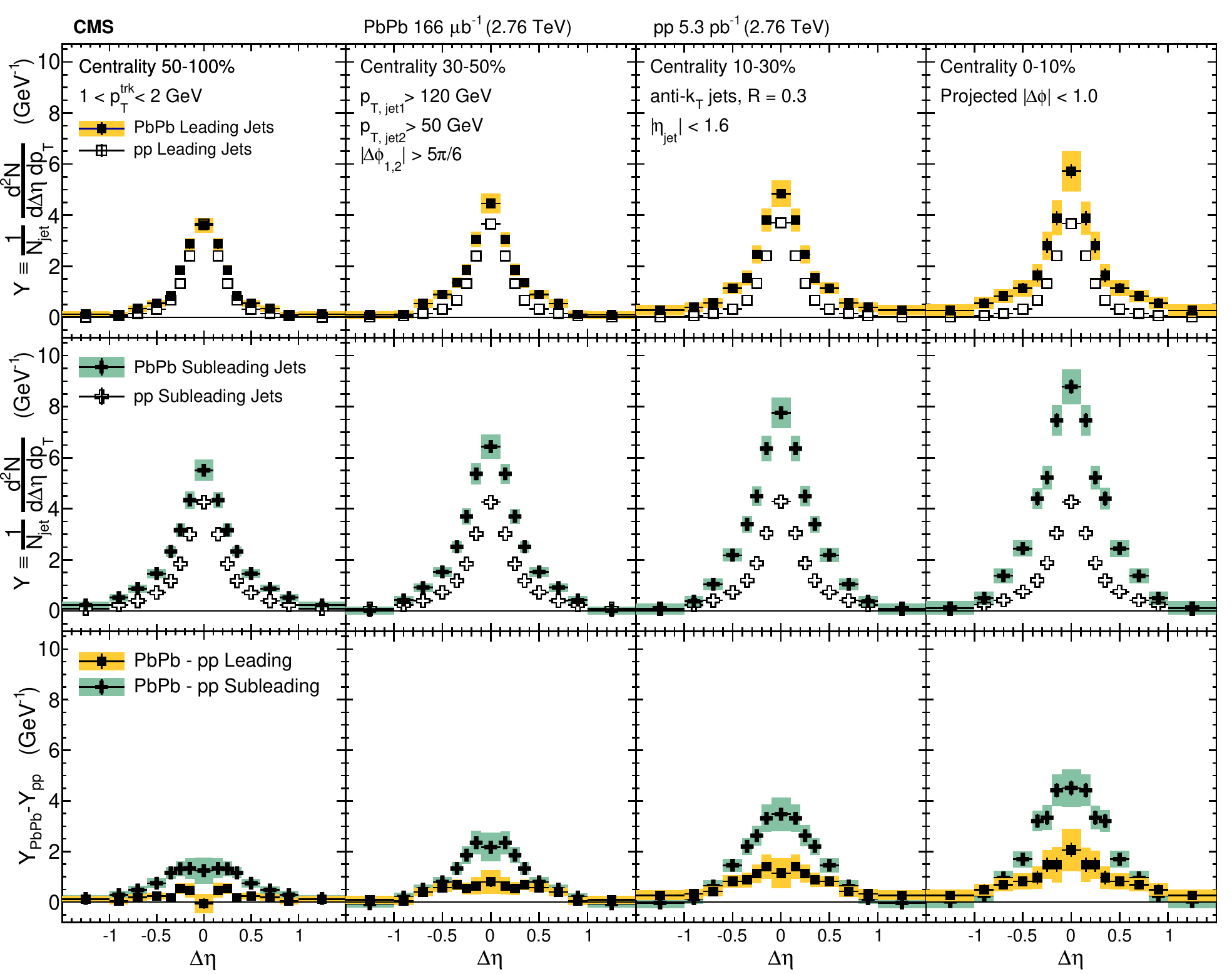}
\caption{The top panels show the $\Delta\eta$ distributions (projected over $\abs{\Delta\phi} < 1$) of charged-particle background-subtracted yields correlated to PbPb and pp leading jets with $p_{\mathrm{T},\text{jet1}}>$ 120\GeV.  The middle panels show the same distributions for subleading jets with $p_{\mathrm{T},\text{jet2}}>$ 50\GeV, and the bottom panels show the difference PbPb minus pp for both leading and subleading jets. The total systematic uncertainties are shown as shaded boxes, and statistical uncertainties are shown as vertical bars (often smaller than the symbol size).}
 \label{fig:fig2_dEta1}
  \end{figure}

\begin{figure}[hbtp]
\centering
\includegraphics[width=0.99\textwidth]{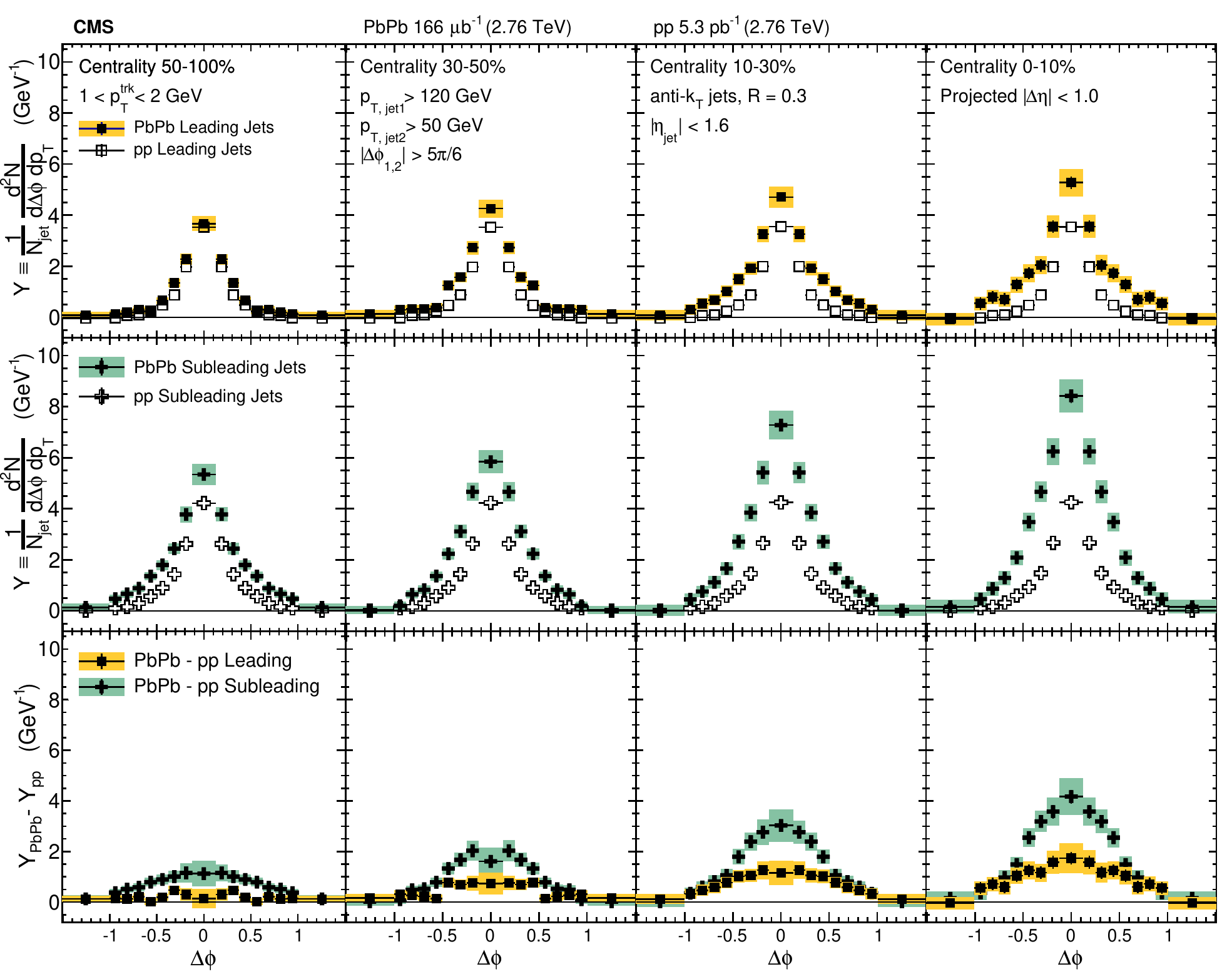}
     \caption{The top panels show the $\Delta\phi$ distributions (projected over $|\Delta\eta < 1$) of charged-particle background-subtracted yields correlated to PbPb and pp leading jets with $p_{\mathrm{T},\text{jet1}}>$ 120\GeV.  The middle panels show the same distributions for subleading jets with $p_{\mathrm{T},\text{jet2}}>$ 50\GeV, and the bottom panels show the difference PbPb minus pp for both leading and subleading jets. The total systematic uncertainties are shown as shaded boxes, and statistical uncertainties are shown as vertical bars (often smaller than the symbol size).}
      \label{fig:fig2_dPhi1}
  \end{figure}

The next two figures present the results of the jet-track correlation measurements for dijets with leading jet $\pt >$ 120\GeV and subleading jet $\pt >$ 50\GeV, obeying the back-to-back angular selection criteria previously described. Figure~\ref{fig:fig2_dEta1} presents the projection of jet-track correlations measured for charged tracks with $\pt^\text{trk}$ between 1 and 2\GeV on the $\Delta\eta$ axis for the leading (upper panel) and subleading (middle panel) jets, while Fig.~\ref{fig:fig2_dPhi1} shows the corresponding projections on the $\Delta\phi$ axis. Again pp data are included for comparison, and for the most peripheral (50--100\% central) PbPb events the correlations are similar to the pp reference for the leading jets, and differ only slightly for the subleading jets. As in the case of inclusive jets, differences of correlations between pp and PbPb collisions gradually increase from peripheral to central collisions, and are most pronounced in the 0--10\% central events for both leading and subleading jets.  We note that there is little difference between the leading and inclusive jet correlated-yield distributions, indicating that the requirement that leading jets have the highest $\pt$ in the event does not significantly bias the selection of jets with $\pt > 120$\GeV.

For this lowest $\pt^\text{trk}$ bin shown, we observe that (as for the inclusive jet selection) the excess of correlated yield extends significantly beyond the typical jet reconstruction radius for both leading and subleading jets.  The soft excess is more pronounced on the more ``quenched'' subleading side, but is also present on the leading side.  This indicates that leading jets, although surface-biased toward shorter path-lengths through the medium, also experience quenching in central PbPb collisions.  To better illustrate both subleading and leading modifications, the last row of Figs.~\ref{fig:fig2_dEta1} and \ref{fig:fig2_dPhi1} shows the differences (PbPb minus pp) of the correlations in the two upper panels.

\begin{figure}[hbtp]
\centering
\includegraphics[width=0.99\textwidth]{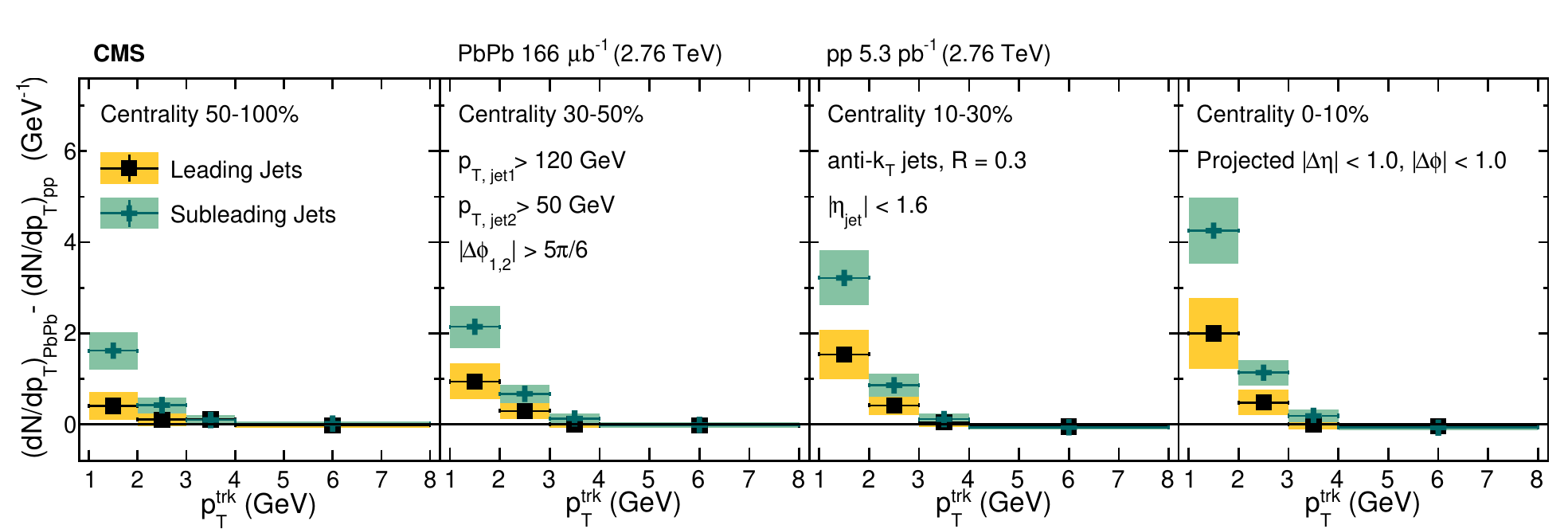}
\caption{Total excess correlated yield observed in the PbPb data with respect to the reference measured in pp collisions, shown as a function of track $\pt$ in four different centrality intervals (0--10\%, 10--30\%, 30--50\%, 50--100\%) for both leading jets with $p_{\mathrm{T},\text{jet1}}> $120\GeV and subleading jets with $p_{\mathrm{T},\text{jet2}}>$ 50\GeV. The total systematic uncertainties are shown as shaded boxes, and statistical uncertainties are shown as vertical bars (often smaller than the symbol size).}
\label{fig:Fig2_Excess_vs_trkpt}
\end{figure}

To quantify the total per-jet excess yield observed in the PbPb data with respect to the pp reference, we plot the integrals of the excess yields (PbPb minus pp) as a function of $\pt^\text{trk}$ and collision centrality in Fig.~\ref{fig:Fig2_Excess_vs_trkpt}.  As the figure shows, in both leading and subleading jets, the excess yield diminishes for higher momentum tracks until the yield becomes similar to the pp reference for the highest $\pt^\text{trk}$ bin of 4--8\GeV.  As seen in previous figures, central collisions exhibit the largest low-$\pt^\text{trk}$ excesses. This demonstrates the expected trend corresponding to quenching of both the leading and the subleading jets, as energy from particles with higher $\pt^\text{trk}$ is redistributed into particles with lower $\pt^\text{trk}$ via interactions with the medium.

\begin{figure}[hbtp]
\centering
\includegraphics[width=0.99\textwidth]{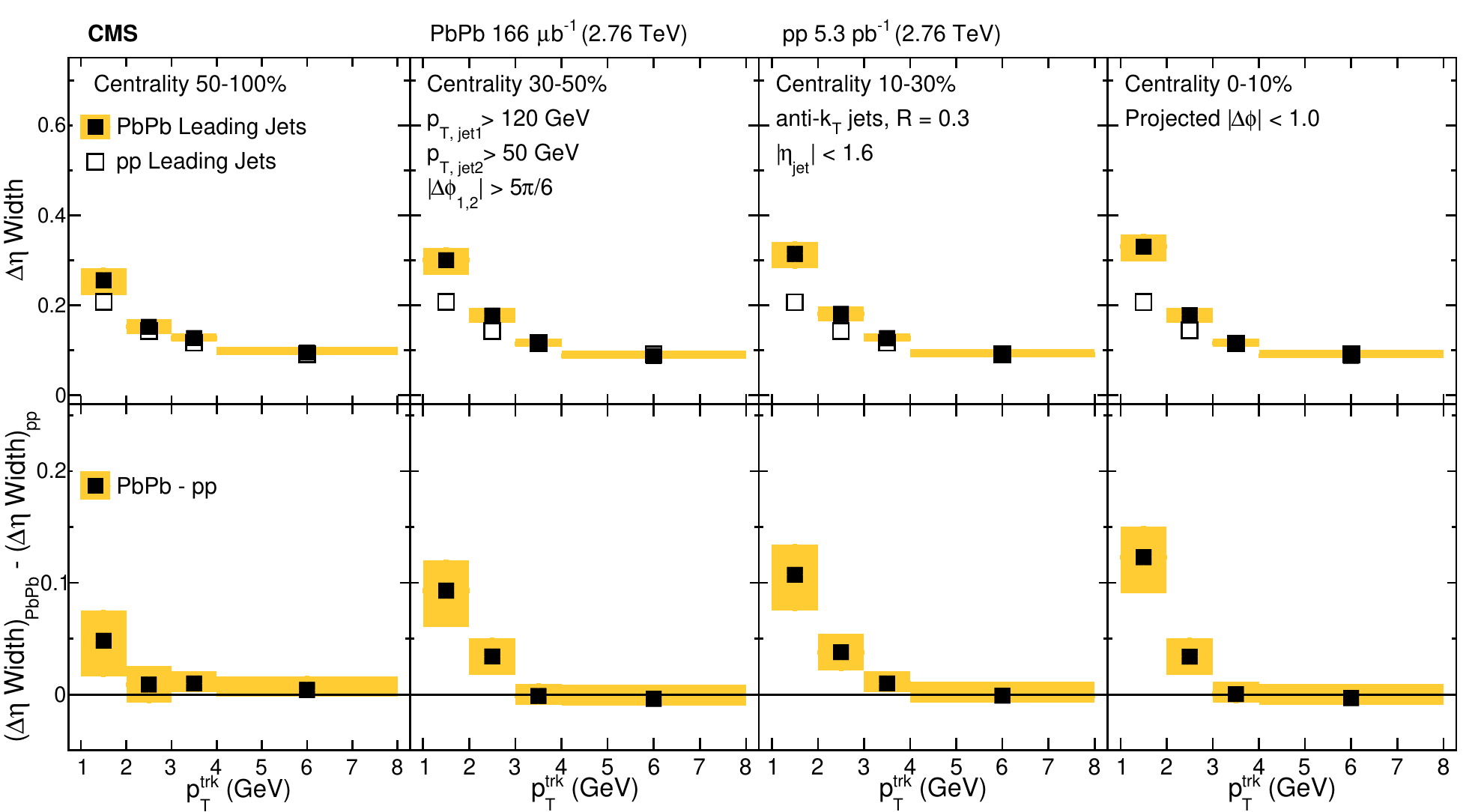}
\caption{Comparison of the widths in PbPb and pp of the $\Delta\eta$ charged-particle distributions correlated to leading jets with $p_{\mathrm{T},\text{jet1}}>$ 120\GeV, as a function of $\pt^\text{trk}$.  The bottom row shows the difference of the widths in PbPb and pp data.  The shaded band corresponds to systematic uncertainty, and statistical uncertainties are smaller than symbol size.}
\label{fig:Width_vs_trkpt}
\end{figure}

\begin{figure}[hbtp]
\centering
\includegraphics[width=0.99\textwidth]{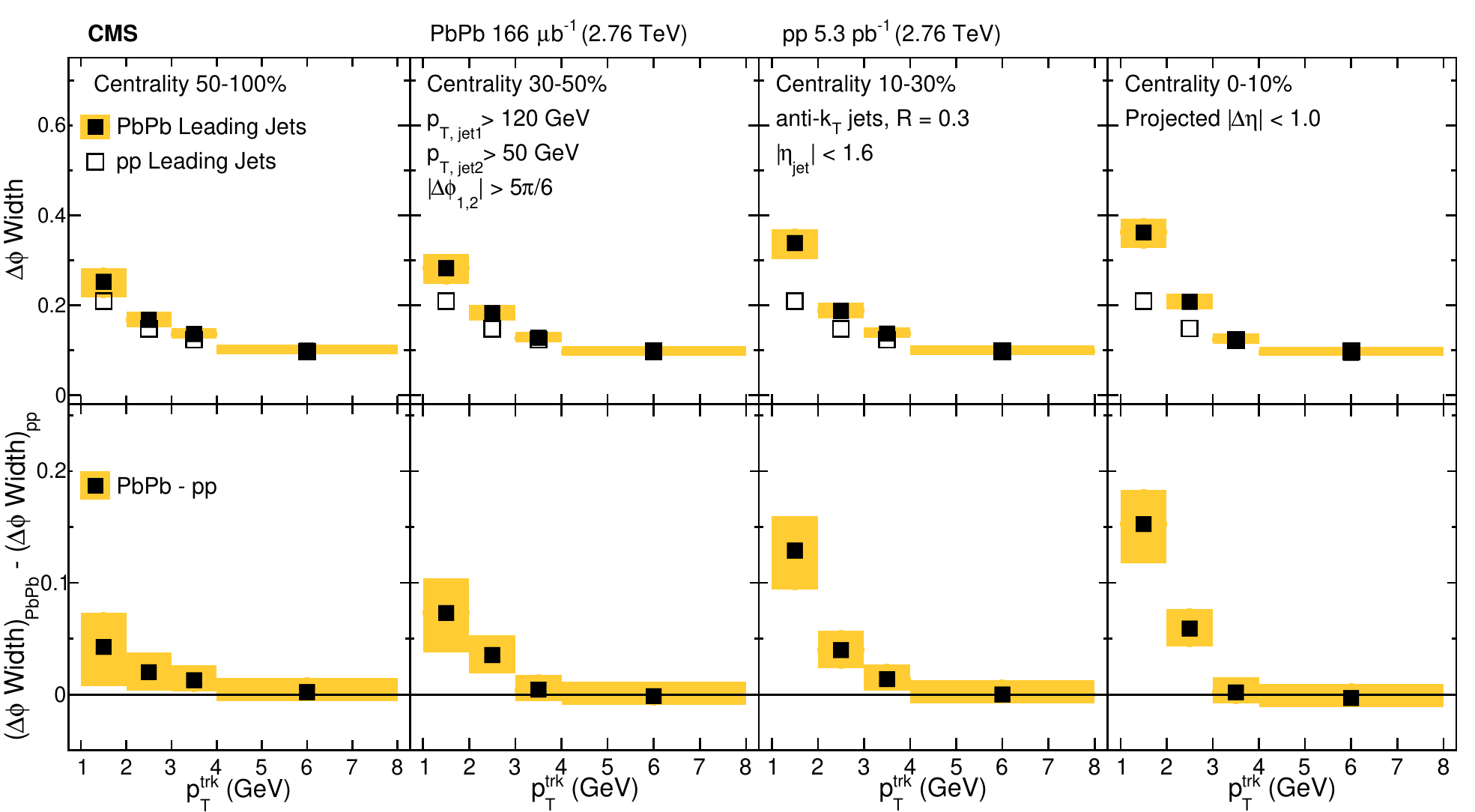}
\caption{Comparison of the widths in PbPb and pp of the $\Delta\phi$ charged-particle distributions correlated to leading jets with $p_{\mathrm{T},\text{jet1}}>$ 120\GeV, as a function of $\pt^\text{trk}$.  The bottom row shows the difference of the widths in PbPb and pp data.  The shaded band corresponds to systematic uncertainty, and statistical uncertainties are smaller than symbol size.}
\label{fig:dphi_Width_vs_trkpt}
\end{figure}

\begin{figure}[hbtp]
\centering
\includegraphics[width=0.99\textwidth]{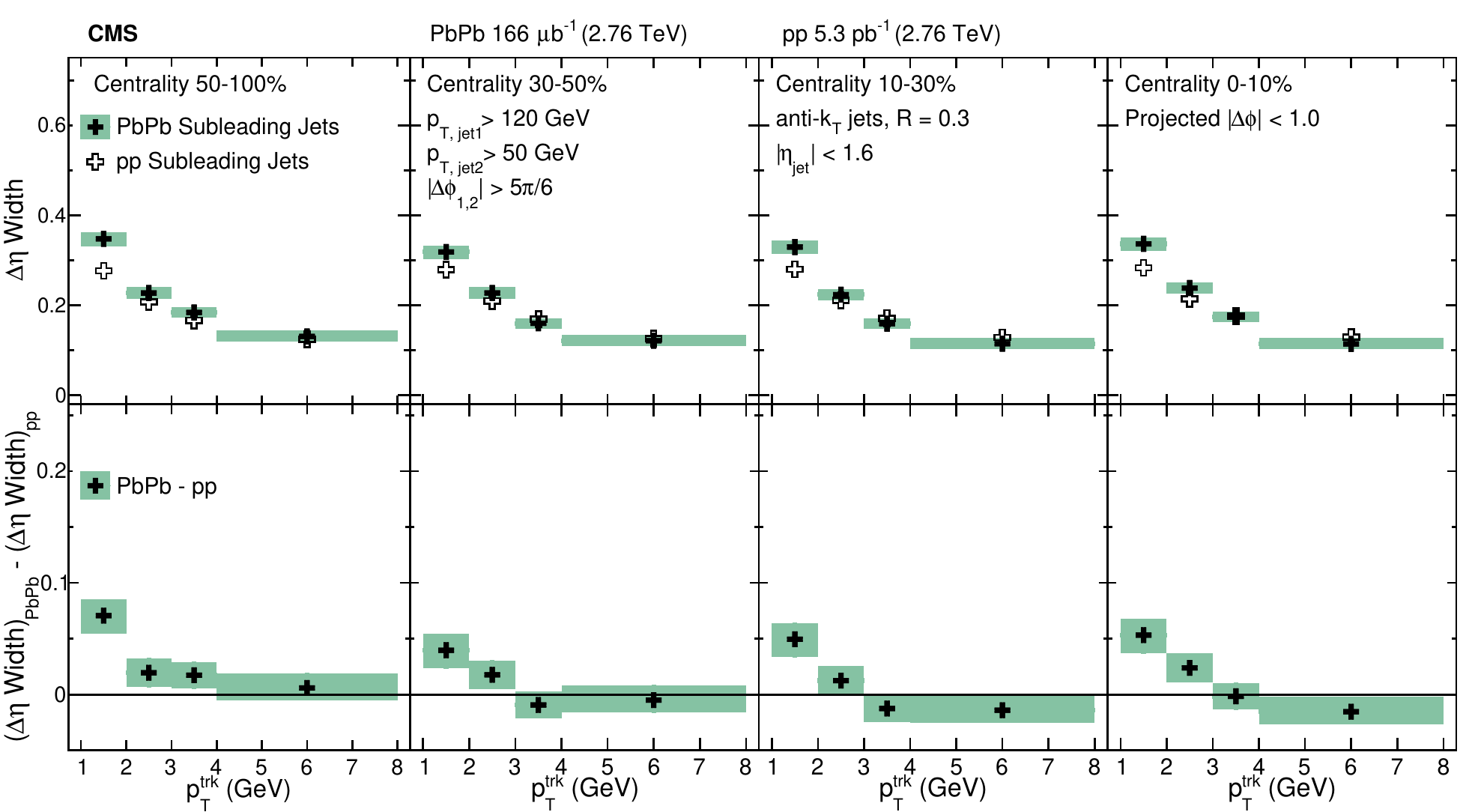}
\caption{Comparison of the widths in PbPb and pp of the $\Delta\eta$ charged-particle distributions correlated to leading jets with $p_{\mathrm{T},\text{jet2}}>$ 50\GeV, as a function of $\pt^\text{trk}$.  The bottom row shows the difference of the widths in PbPb and pp data.  The shaded band corresponds to systematic uncertainty, and statistical uncertainties are smaller than symbol size.}
\label{fig:Width_vs_trkpt_sub}
\end{figure}

\begin{figure}[hbtp]
\centering
\includegraphics[width=0.99\textwidth]{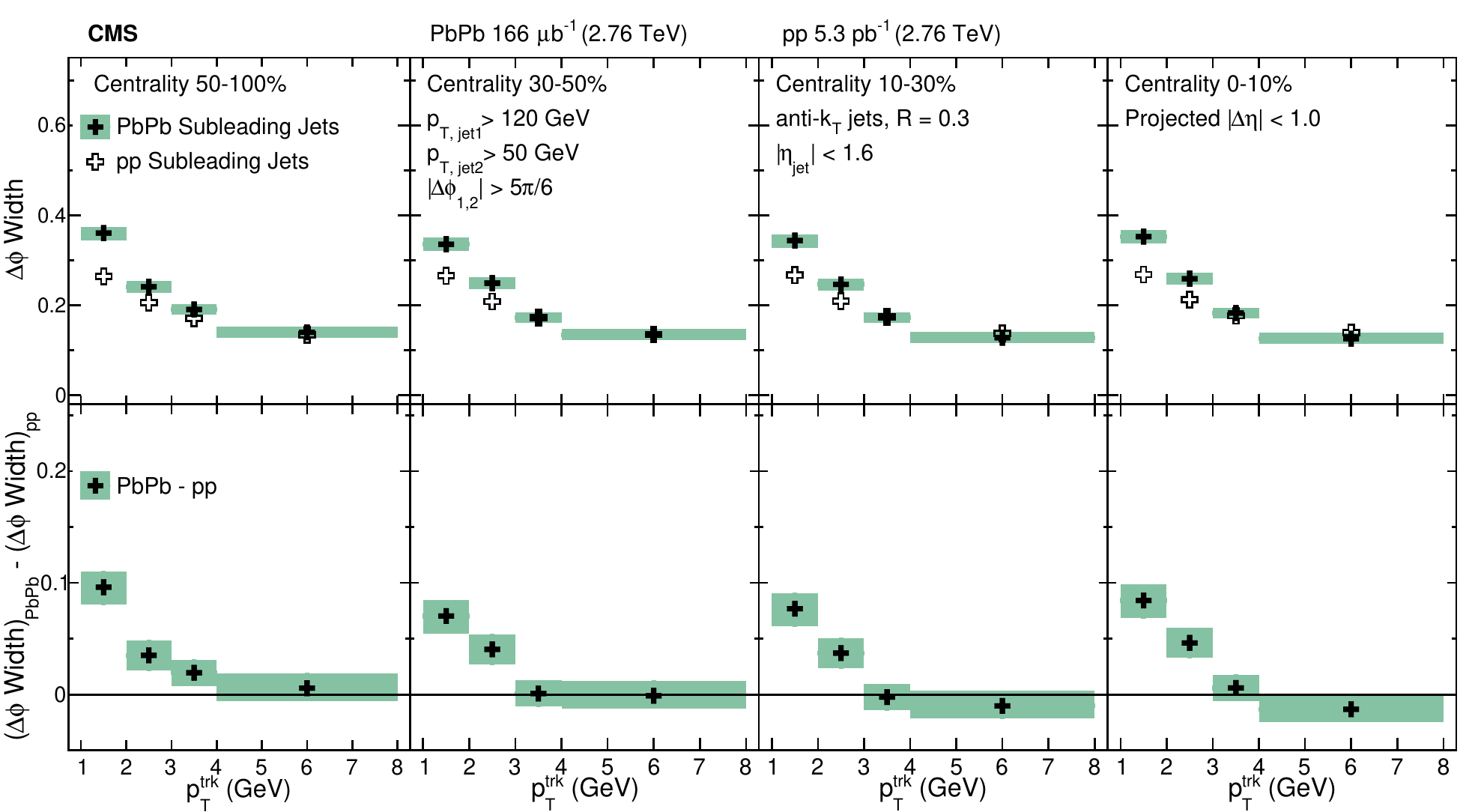}
\caption{Comparison of the widths in PbPb and pp of the $\Delta\phi$ charged-particle distributions correlated to leading jets with $p_{\mathrm{T},\text{jet2}}>$ 50\GeV, as a function of $\pt^\text{trk}$.  The bottom row shows the difference of the widths in PbPb and pp data.  The shaded band corresponds to systematic uncertainty, and statistical uncertainties are smaller than symbol size.}
\label{fig:dphi_Width_vs_trkpt_sub}
\end{figure}

In order to characterize the angular widths of the charged-particle distributions in $\Delta\eta$ and $\Delta\phi$, we fit the measured correlations with a double Gaussian function (which was found to best describe the overall correlation shapes).  The width is defined as the region around zero in $\abs{\Delta\eta}$ or $\abs{\Delta\phi}$ that contains 67\% of the total correlated yield.  Width uncertainties are calculated by repeating the measurement for the $\Delta\eta$ and $\Delta\phi$ distributions varied by their respective systematic uncertainties, which are treated as fully correlated for the purposes of this determination.  Widths for leading and subleading jet correlations in $\Delta\eta$ and $\Delta\phi$ are presented as a function of $\pt^\text{trk}$ in Figs.~\ref{fig:Width_vs_trkpt}--\ref{fig:dphi_Width_vs_trkpt_sub}. Distributions of low-$\pt$ tracks correlated with either of the two jets are found to be significantly broader in central PbPb events compared to those in pp data in both $\Delta\eta$ and $\Delta\phi$ dimensions. This broadening is greatest for the low-$\pt$ tracks and in the most central events, and diminishes quickly with increasing track momenta.  Above 4\GeV, the widths measured in PbPb and pp events are the same within the systematic uncertainties.  We note that the width of the PbPb minus pp excess yield is similar for leading and subleading jets.  In pp data, however, the peak associated with the subleading jet is softer and broader than the peak associated with the leading jet.  There is therefore a larger difference in peak width when comparing PbPb leading jet peaks to the narrow pp leading jet peaks (Figs.~\ref{fig:Width_vs_trkpt}--\ref{fig:dphi_Width_vs_trkpt}), and a smaller difference when comparing PbPb subleading jet peaks to the broader pp subleading jet peaks (Figs.~\ref{fig:Width_vs_trkpt_sub}--\ref{fig:dphi_Width_vs_trkpt_sub}).

\section{Summary}
In this analysis, jet-track correlations have been studied as a function of $\Delta\eta$ and $\Delta\phi$ with respect to the jet axis in PbPb and pp collisions at $\sqrtsNN = 2.76$~TeV.  Two-dimensional angular correlations have been considered for charged particles with $\pt^\text{trk}>1$\GeV as a function of $\pt^\text{trk}$ and collision centrality for two jet selections.  A sample of inclusive jets above the jet momentum threshold of 120\GeV was studied, as well as a sample of dijet events selected to include a leading jet with $\pt>120$\GeV and a subleading jet with $\pt>50$\GeV. In all cases, an excess of soft particle yields was observed in central PbPb collisions with respect to pp reference data, similar for inclusive and leading jet samples and larger for the (more-quenched) subleading jet sample.  The low-$\pt^\text{trk}$ (1--3\GeV) excess-yield distributions were studied individually and, in both $\Delta\eta$ and $\Delta\phi$, they exhibit similar Gaussian-like distributions out to large relative angles ($\Delta\eta\approx 1$ and $\Delta\phi\approx 1$) from the jet axis.  The excess was found to be largest at the lowest $\pt^\text{trk}$ (1--2\GeV) in the most central (0--10\%) PbPb data, and to decrease gradually with centrality.  For peripheral (50--100\%) PbPb collisions, correlated low-$\pt^\text{trk}$ particle yields are only slightly larger than those for the pp reference.  The excess also gradually decreases with increasing $\pt^\text{trk}$ until yields of particles with $\pt^\text{trk}>4$\GeV are similar to pp reference data, consistent with the results of a previous CMS jet quenching study.  This new correlation analysis provides a comprehensive evaluation of medium effects on jet properties, extending information about jet shapes to large angles away from the jet axis.

\begin{acknowledgments}
We congratulate our colleagues in the CERN accelerator departments for the excellent performance of the LHC and thank the technical and administrative staffs at CERN and at other CMS institutes for their contributions to the success of the CMS effort. In addition, we gratefully acknowledge the computing centers and personnel of the Worldwide LHC Computing Grid for delivering so effectively the computing infrastructure essential to our analyses. Finally, we acknowledge the enduring support for the construction and operation of the LHC and the CMS detector provided by the following funding agencies: BMWFW and FWF (Austria); FNRS and FWO (Belgium); CNPq, CAPES, FAPERJ, and FAPESP (Brazil); MES (Bulgaria); CERN; CAS, MoST, and NSFC (China); COLCIENCIAS (Colombia); MSES and CSF (Croatia); RPF (Cyprus); MoER, ERC IUT and ERDF (Estonia); Academy of Finland, MEC, and HIP (Finland); CEA and CNRS/IN2P3 (France); BMBF, DFG, and HGF (Germany); GSRT (Greece); OTKA and NIH (Hungary); DAE and DST (India); IPM (Iran); SFI (Ireland); INFN (Italy); MSIP and NRF (Republic of Korea); LAS (Lithuania); MOE and UM (Malaysia); CINVESTAV, CONACYT, SEP, and UASLP-FAI (Mexico); MBIE (New Zealand); PAEC (Pakistan); MSHE and NSC (Poland); FCT (Portugal); JINR (Dubna); MON, RosAtom, RAS and RFBR (Russia); MESTD (Serbia); SEIDI and CPAN (Spain); Swiss Funding Agencies (Switzerland); MST (Taipei); ThEPCenter, IPST, STAR and NSTDA (Thailand); TUBITAK and TAEK (Turkey); NASU and SFFR (Ukraine); STFC (United Kingdom); DOE and NSF (USA).

Individuals have received support from the Marie-Curie program and the European Research Council and EPLANET (European Union); the Leventis Foundation; the A. P. Sloan Foundation; the Alexander von Humboldt Foundation; the Belgian Federal Science Policy Office; the Fonds pour la Formation \`a la Recherche dans l'Industrie et dans l'Agriculture (FRIA-Belgium); the Agentschap voor Innovatie door Wetenschap en Technologie (IWT-Belgium); the Ministry of Education, Youth and Sports (MEYS) of the Czech Republic; the Council of Science and Industrial Research, India; the HOMING PLUS program of the Foundation for Polish Science, cofinanced from European Union, Regional Development Fund; the OPUS program of the National Science Center (Poland); the Compagnia di San Paolo (Torino); MIUR project 20108T4XTM (Italy); the Thalis and Aristeia programs cofinanced by EU-ESF and the Greek NSRF; the National Priorities Research Program by Qatar National Research Fund; the Rachadapisek Sompot Fund for Postdoctoral Fellowship, Chulalongkorn University (Thailand); the Chulalongkorn Academic into Its 2nd Century Project Advancement Project (Thailand); and the Welch Foundation, contract C-1845.
\end{acknowledgments}
\bibliography{auto_generated}
\cleardoublepage \appendix\section{The CMS Collaboration \label{app:collab}}\begin{sloppypar}\hyphenpenalty=5000\widowpenalty=500\clubpenalty=5000\textbf{Yerevan Physics Institute,  Yerevan,  Armenia}\\*[0pt]
V.~Khachatryan, A.M.~Sirunyan, A.~Tumasyan
\vskip\cmsinstskip
\textbf{Institut f\"{u}r Hochenergiephysik der OeAW,  Wien,  Austria}\\*[0pt]
W.~Adam, E.~Asilar, T.~Bergauer, J.~Brandstetter, E.~Brondolin, M.~Dragicevic, J.~Er\"{o}, M.~Flechl, M.~Friedl, R.~Fr\"{u}hwirth\cmsAuthorMark{1}, V.M.~Ghete, C.~Hartl, N.~H\"{o}rmann, J.~Hrubec, M.~Jeitler\cmsAuthorMark{1}, V.~Kn\"{u}nz, A.~K\"{o}nig, M.~Krammer\cmsAuthorMark{1}, I.~Kr\"{a}tschmer, D.~Liko, T.~Matsushita, I.~Mikulec, D.~Rabady\cmsAuthorMark{2}, N.~Rad, B.~Rahbaran, H.~Rohringer, J.~Schieck\cmsAuthorMark{1}, R.~Sch\"{o}fbeck, J.~Strauss, W.~Treberer-Treberspurg, W.~Waltenberger, C.-E.~Wulz\cmsAuthorMark{1}
% ...
\vskip\cmsinstskip
% ...
1:~~Also at Vienna University of Technology, Vienna, Austria\\
2:~~Also at CERN, European Organization for Nuclear Research, Geneva, Switzerland\\
% ...

\end{sloppypar}
\end{document}